\documentclass[aps, reprint, twocolumn,superscriptaddress]{revtex4-1}
\usepackage[latin1]{inputenc}
\usepackage{graphicx}
\usepackage{bm}
\usepackage{amsmath,amssymb,amstext,graphicx,bm,booktabs}
\newcommand{\Cshunt}{\ensuremath{C_\mathrm{M}} }
\newcommand{\Cseries}{\ensuremath{C_\mathrm{D}} }
\newcommand{\CS}{\ensuremath{C_\mathrm{S}} }

\newcommand{\VS}{\ensuremath{V_{\text{S}}}}
\newcommand{\Vshunt}{\ensuremath{V_{\text{M}}}}
\newcommand{\VM}{\ensuremath{V_{\text{m}}}}
\newcommand{\VD}{\ensuremath{V_{\text{D}}}}
\newcommand{\VL}{\ensuremath{V_{\text{L}}}}
\newcommand{\Vbias}{\ensuremath{V_{\text{bias}}}}

\newcommand{\fC}{\ensuremath{f_\mathrm{C}} }
\newcommand{\fM}{\ensuremath{f_\mathrm{m}} }
\newcommand{\fo}{\ensuremath{f_0} }

\newcommand{\SC}{\ensuremath{S_C}}
\newcommand{\deltaC}{\ensuremath{\delta C}}

\newcommand{\TMC}{\ensuremath{T_\mathrm{MC}}}

\newcommand{\ZO}{\ensuremath{Z_0} }

\newcommand{\Cdot}{\ensuremath{C_\mathrm{dot}} }

\newcommand{\GdotDC}{\ensuremath{G_\mathrm{dot}^\mathrm{DC}} }

\begin{document}

\title{Sensitive radio-frequency measurements of a quantum dot by tuning to perfect impedance matching}

\author{N.~Ares}
\affiliation{Department of Materials, University of Oxford, Parks Road, Oxford OX1 3PH, United Kingdom}

\author{F.J.~Schupp }
\affiliation{Department of Materials, University of Oxford, Parks Road, Oxford OX1 3PH, United Kingdom}

\author{A.~Mavalankar}
\affiliation{Department of Materials, University of Oxford, Parks Road, Oxford OX1 3PH, United Kingdom}

\author{G.~Rogers}
\affiliation{Department of Materials, University of Oxford, Parks Road, Oxford OX1 3PH, United Kingdom}

\author{J.~Griffiths}
\affiliation{Cavendish Laboratory, J. J. Thomson Avenue, Cambridge CB3 0HE, United Kingdom}

\author{G.A.C.~Jones}
\affiliation{Cavendish Laboratory, J. J. Thomson Avenue, Cambridge CB3 0HE, United Kingdom}

\author{I.~Farrer}
\affiliation{Cavendish Laboratory, J. J. Thomson Avenue, Cambridge CB3 0HE, United Kingdom}

\author{D.A.~Ritchie}
\affiliation{Cavendish Laboratory, J. J. Thomson Avenue, Cambridge CB3 0HE, United Kingdom}

\author{C.G.~Smith}
\affiliation{Cavendish Laboratory, J. J. Thomson Avenue, Cambridge CB3 0HE, United Kingdom}

\author{A.~Cottet}
\affiliation{Laboratoire Pierre Aigrain, Ecole Normale Sup\'{e}rieure-PSL Research University,
CNRS, Universit\'{e} Pierre et Marie Curie-Sorbonne Universit\'{e}s,
Universit\'{e} Paris Diderot-Sorbonne Paris Cit\'{e},
24 rue Lhomond, 75231 Paris Cedex 05, France}

\author{G.A.D.~Briggs}
\affiliation{Department of Materials, University of Oxford, Parks Road, Oxford OX1 3PH, United Kingdom}

\author{E.A.~Laird}
\affiliation{Department of Materials, University of Oxford, Parks Road, Oxford OX1 3PH, United Kingdom}

\begin{abstract}

Electrical readout of spin qubits requires fast and sensitive measurements, which are hindered by poor impedance matching to the device. We demonstrate perfect impedance matching in a radio-frequency readout circuit, using voltage-tunable varactors to cancel out parasitic capacitances. An optimized capacitance sensitivity of $1.6~\mathrm{aF}/\sqrt{\mathrm{Hz}}$ is achieved at a maximum source-drain bias of $170~\mu$V root-mean-square and with a bandwidth of 18~MHz. Quantum dot Coulomb blockade is measured in both conductance and capacitance, and the two contributions are found to be proportional as expected from a quasistatic tunneling model. 
We benchmark our results against the requirements for single-shot qubit readout using quantum capacitance, a goal that has so far been elusive.

\end{abstract}

\date{\today{}}
\maketitle

\section{Introduction}
Measuring the quantum state of an electronic device with high fidelity requires sensitive, fast, and non-invasive readout. If the state can be mapped to an electrical impedance, this can be achieved using radio-frequency reflectometry of an electrical resonator incorporating the quantum device~\cite{Schoelkopf_1998}. This permits rapid readout of charge sensors~\cite{Cassidy_2007, Reilly_2007}, spin qubits~\cite{Barthel_2010}, and nanomechanical resonators~\cite{LaHaye_2009}, as well as complex impedance measurements of quantum dot circuits~\cite{Petersson_2010, Chorley_2012, Jung_2012, Schroer_2012, Colless_2013, Gonzalez_2015}. For optimal sensitivity, which can approach the quantum limit~\cite{Xue2009}, impedance matching between the device and the external circuitry is essential to maximize power transfer between them~\cite{Aassime_2001}. This is made challenging by the large resistance typical of quantum dot devices ($\gtrsim$100~k$\Omega$, compared with usual line impedance $\ZO=50\Omega$), and by parasitic capacitances in the matching circuit which vary unpredictably between devices.

We present a circuit that achieves controllable perfect matching with a high device impedance, even accounting for parasitics. Voltage-tunable capacitors allow \emph{in situ} tuning of the matching condition~\cite{Muller_2010, Hellmuller_2012} and an absolute calibration of the capacitance sensitivity. We measure the complex impedance of a Coulomb-blockaded quantum dot, finding that the capacitance changes in proportion to the conductance. This relation is in agreement with a quasi-static model of electron tunneling.


\begin{figure*}
\includegraphics{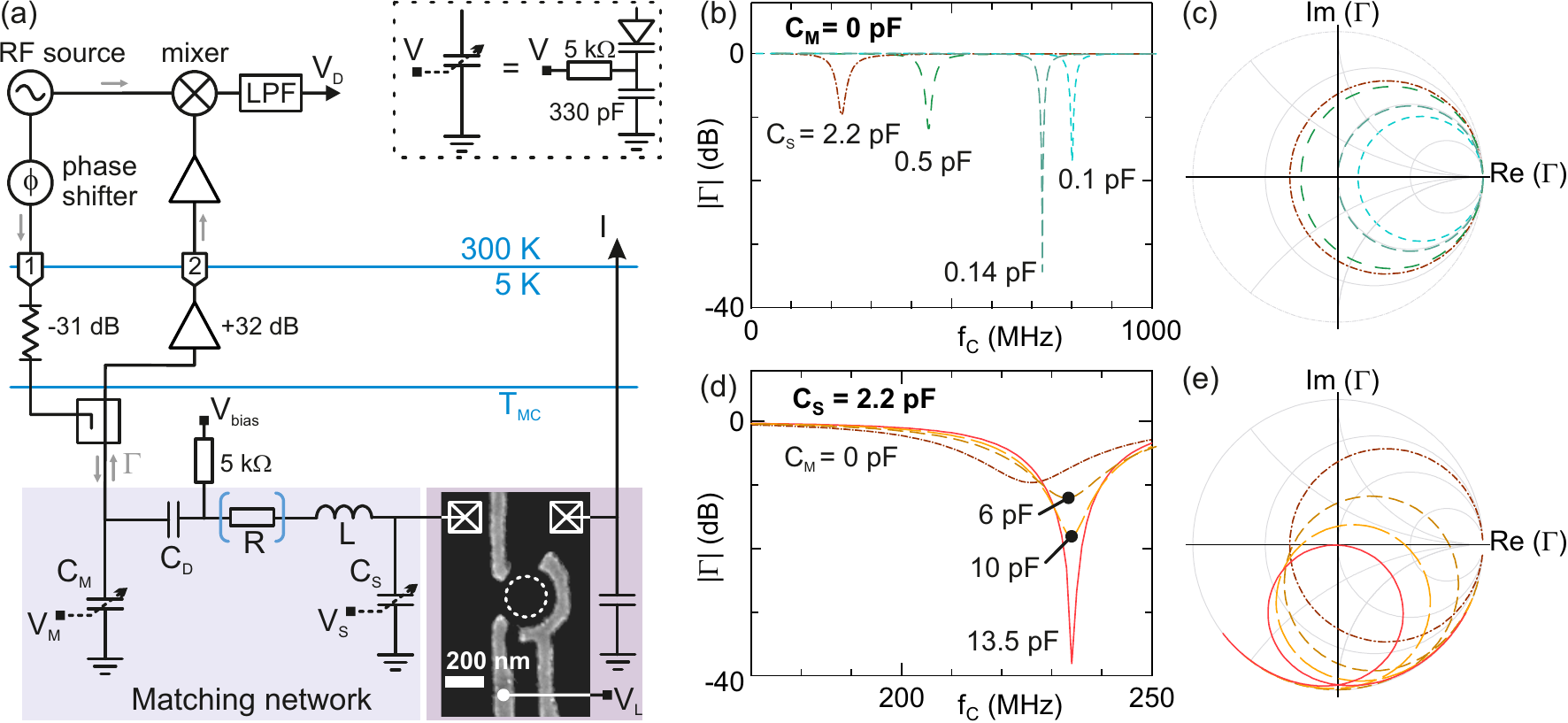}
\caption{\label{Fig1}
(a)
Experimental setup. A gate-defined quantum dot (electron micrograph right, with ohmic contacts denoted by boxes) is coupled to an impedance matching network formed from an inductor $L$ (223~nH), variable capacitors \CS and $\Cshunt$ (tuned through the circuit shown in the inset), and fixed capacitor \Cseries (87~pF). Parasitic losses in the circuit are parameterized by an effective resistance $R$. To probe the matching network, a radio-frequency signal is injected at port 1, passed via a directional coupler, and after reflection and amplification received at port 2. The reflected signal is demodulated at room temperature to a DC voltage $\VD$ by mixing it with a local oscillator; by adjusting the phase shift $\phi$, different quadratures of the signal can be detected. Alternatively, the signal is measured using a network analyser or spectrum analyser. A bias resistor allows measurements of the device current $I$ with DC bias $\Vbias$. 
(b), (c)
Simulation with no matching capacitor ($\Cshunt=0$). Voltage reflection coefficient $\Gamma$ is plotted as a function of frequency for different values of sample capacitance $\CS$, as magnitude (b) and as a Smith chart~\cite{Pozar}~(c). The effective resistance is taken as $R=20~\Omega$, the device resistance as 1~G$\Omega$, and specified non-idealities of the inductor are included (see Supplementary Information). The capacitance of the device is taken as included in $\CS$. 
Perfect matching occurs when $\Gamma$ crosses the origin of the Smith chart ($|\Gamma|=0$). With these parameters, this is achieved only when $\CS=0.14$~pF, less than typical parasitic values. (d), (e) Simulated reflection for varying $\Cshunt$. Perfect matching can be achieved even for a realistic large value of $\CS$ (here at $\Cshunt=13.5$~pF for $\CS=2.2$~pF). In (c) and (e), grey contours on the Smith chart indicate constant real or imaginary circuit input impedance.
}
\end{figure*}

\section{Reflectometry with perfect impedance matching}
We implement the matching scheme using as the device under test a gate-defined GaAs quantum dot, measured in a dilution refrigerator as shown in Fig.~\ref{Fig1}(a). The impedance matching network is realized with a chip inductor $L$ and capacitors $\Cshunt, \Cseries, \CS$, forming a resonant circuit incorporating the device. To make a reflectometry measurement, a radio-frequency signal with frequency $\fC$, injected at port 1 of the cryostat, is coupled via a directional coupler to the matching network input. The reflected amplified signal is returned to port~2. From the amplitude and phase of this signal, the reflection coefficient of the resonant circuit and therefore the complex impedance of the device can be deduced. A room-temperature homodyne detection circuit demodulates a chosen quadrature of the reflected signal to a DC signal $V_\mathrm{D}$. Simultaneous DC transport measurements are made using a tee to apply a bias voltage $V_\mathrm{bias}$.

In previous work~\cite{Schoelkopf_1998,Reilly_2007,Cassidy_2007,Petersson_2010,Chorley_2012,Muller_2010}, the impedance match is usually hindered by parasitic capacitances. Even with careful engineering, sample wiring typically contributes a sample capacitance $\CS\gtrsim 0.3$~pF in parallel with the device~\cite{Hornibrook_2014}. In our experiment, these parasitic capacitances are mitigated by adding a matching capacitors $\Cshunt$ and a decoupling capacitor $\Cseries$ at the input of the matching network. This is illustrated in~Fig.~\ref{Fig1}(b) to (e), which show simulated reflection coefficient $\Gamma$ as a function of frequency for typical device parameters. With no matching capacitor (Fig.~\ref{Fig1}(b) and (c)), perfect matching (indicated by zero reflection) is achieved only for one value of $\CS$, in this case 0.14~pF. With a parasitic capacitance above this value, perfect matching cannot be achieved at any carrier frequency $\fC$, degrading the sensitivity. One approach to restore matching is to increase the inductance $L$; however, this reduces the readout bandwidth, and more problematically introduces self-resonances of the inductor close to the operating frequency. Our approach is to introduce the capacitor $\Cshunt$ to cancel out a reactive contribution to the impedance. By increasing $\Cshunt$, a perfect match can be achieved even with a much larger value of $\CS$ (Fig.~\ref{Fig1}(d) and (e)). In this scheme, the purpose of $\Cseries$ is to increase slightly the quality factor $Q$ of the circuit by decoupling it from the input.

\begin{figure}
\includegraphics[width=\columnwidth]{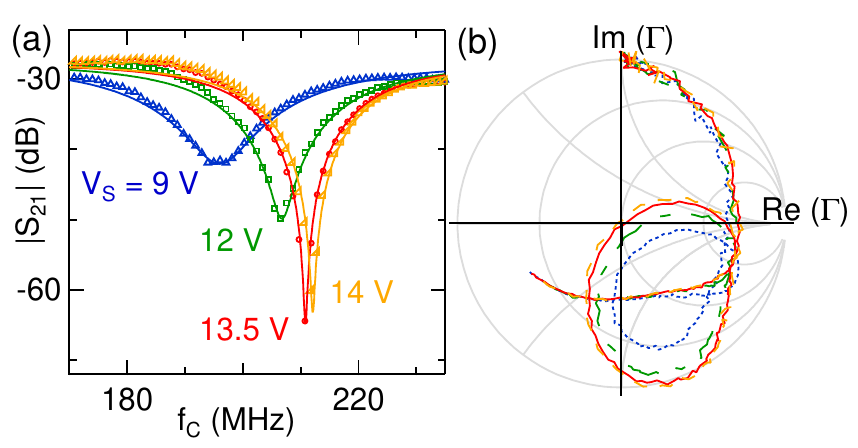}
\caption{\label{Fig2}
Reflectometry measurements as a function of frequency for different $\VS$ settings. Data is taken at $\TMC=1$~K with $\Vshunt$ held constant. (a) Magnitude of $S_{21}$, with data (symbols) fitted with a circuit model (lines). (b) The same data, converted to circuit reflectance and plotted on a Smith chart. Perfect matching is achieved for $\VS=13.5$~V. The direction in which the traces cross the origin of the Smith chart is opposite to Fig.~\ref{Fig1}(c) because the effective resistance $R$ also changes with $\VS$.
}
\end{figure}

This scheme is implemented using varactors (Macom MA46H204-1056) for $\Cshunt$ and $\CS$, controlled by voltages $\Vshunt$ and $\VS$, so that the parameters of the matching network can be tuned \emph{in situ}~(Fig.~\ref{Fig2}). The device under test is a laterally defined quantum dot~\cite{Mavalankar_2013}, fabricated by patterning Ti/Au gates over a GaAs/AlGaAs heterostructure containing a two-dimensional electron gas (depth 90~nm, mobility 125 m$^2$V$^{-1}$s$^{-1}$, carrier concentration $1.31\times 10^{15}$ m$^{-2}$). The device chip was bonded to a printed circuit board mounted with components of the matching circuit. Bias voltages for the varactors and the quantum dot gates were applied through filtered wires with a bandwidth $\sim 100$~kHz. A bias tee (not shown) allowed a high-frequency signal to be added to $\VL$ for characterization at higher frequency.

At a refrigerator temperature $\TMC = 1$~K, gate voltages were set to pinch off the quantum dot completely (device resistance $>200~\mathrm{M}\Omega$). The quality of the impedance match in this configuration was probed by measuring the transmission $S_{21}$ between ports 1 and 2, which is proportional to $\Gamma$. With $\Cshunt$ set to the upper end of its range ($\Cshunt \sim 14$ pF), Fig.~\ref{Fig2}(a) shows $S_{21}$ as a function of frequency for different values of $\VS$. As $\VS$ is increased, the resonance frequency $\fo$ increases, confirming the change in $\CS$. The quality of the match depends strongly on $\CS$, with a minimum in the reflected power near $\VS = 13.5$~V. From fits to these data using a simple circuit model, parameters can be estimated as follows (see also Supplementary Information): From the trace with $\VS=13.5$~V, the tuned capacitances $\Cshunt$ and $\CS$, the effective resistance $R$ characterizing parasitic losses, and the cable insertion loss can be extracted. Traces for other values of $\VS$ are then well reproduced using only $\CS$ and $R$ as free parameters. Perfect matching is achieved at $\fC\approx 211$~MHz and, according to this model, with $\CS\approx2.78$~pF. Using the inferred insertion loss and the known amplifier gain, which give the proportionality constant between $S_\mathrm{21}$ and $\Gamma$, the complex reflection coefficient $\Gamma$ can be plotted on a Smith chart (Fig.~\ref{Fig2}(b)). As $\VS$ is tuned, the traces cross the origin, confirming that the minimum seen in Fig.~\ref{Fig2}(a) indeed indicates a perfect match.

\section{Characterizing the capacitance sensitivity}
\begin{figure}
\includegraphics[width=\columnwidth]{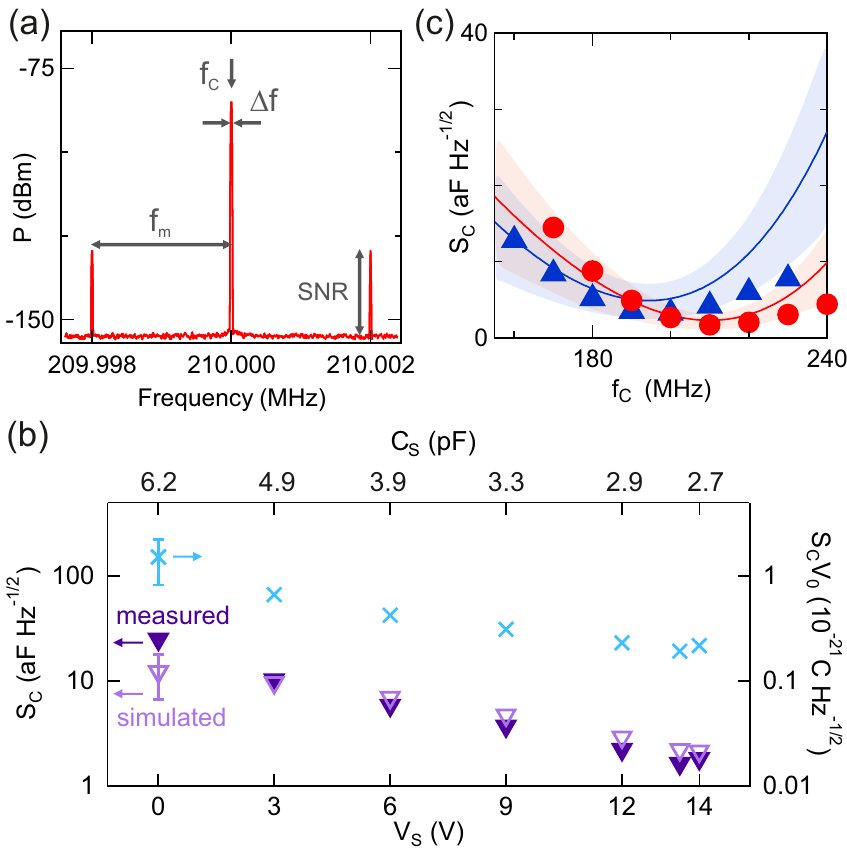}
\caption{
(a) Power spectrum of signal at port 2 near perfect matching ($\VS = 13.5$~V) with varactor modulation  $\VM=2~\mathrm{mV_{RMS}}$, showing the carrier peak and modulation sidebands. The signal-to-noise ratio (SNR) and resolution bandwidth ($\Delta f$) are indicated. (b) Capacitance sensitivity $\SC$ (left axis)  and figure of merit $\SC V_0$ (right axis) as a function of \VS, measured (squares) and simulated (crosses). 
Agreement is good except for $\VS \leq 3$~V, where \fo approaches a resonance of the cryostat. Measurement parameters were $\Delta f=10$~Hz, $\VM=2~\mathrm{mV_{RMS}}$, $\fM=1.75$~kHz and $\fC=\fo$. Tuning the circuit near perfect matching improves the sensitivity to  $\SC=1.6~\mathrm{aF}/\sqrt{\mathrm{Hz}}$. Fitted values of $\CS$ at each $\VS$ setting are marked on the top axis. Error bars on the data are smaller than the symbols; Error bars for $\SC V_0$ and simulated $\SC$ reflect systematic uncertainty in the power delivered to the matching network (see Supplementary Information). For clarity, only a single error bar is marked. 
(c)  Symbols: Measured $\SC$ as a function of $\fC$ for $\VS=13.5$~V (circles) and $\VS=9$~V (triangles). Curves: simulated $\SC$. Shaded bands indicate systematic uncertainty in the simulation of the same origin as in (b). 
\label{Fig3}
}
\end{figure}

The ability to tune the circuit into perfect matching allows for highly sensitive capacitance measurements. This is demonstrated in Fig.~3, which characterizes the sensitivity by detecting the response to a known capacitance change. A sinusoidal signal with root-mean-square (RMS) amplitude $\VM$ and frequency $\fM =$ 1.75~kHz was added to $\VS$ to modulate $\CS$ by a known amount $\deltaC$ (see Supplementary Information). Modulating $\CS$ we guarantee that the response is purely capacitive, unlike modulations on the quantum dot impedance that result in both a capacitive and a resistive response.
As a result of this modulation of $\CS$, the power $P$ detected at port 2 shows sidebands at $\fC \pm \fM$ (Fig.~\ref{Fig3}(a)). From the height of the sidebands above the noise floor, the sensitivity is given by $\SC=\frac{1}{\sqrt{2}}\deltaC (\Delta f)^{-1/2} 10^{-\text{SNR/20}}$, where SNR is the sideband signal-to-noise ratio expressed in dB and $\Delta f$ is the resolution bandwidth~\cite{Brenning_2006}. Over the range of varactor settings measured, $\SC$ is found to change by a factor $>15$, with the best sensitivity close to perfect matching as expected~(Fig.~\ref{Fig3}(b)). This dependence is reproduced well by the same circuit model as above (see Supplementary Information). In agreement with the model, $\SC$ is optimized when \fC is set to the resonance frequency \fo (Fig.~\ref{Fig3}(c)). The optimum sensitivity, attained at $\VS=13.5$~V and  $\fC=\fo=~210.75$~MHz, is $\SC=1.6$~aF$/\sqrt{\text{Hz}}$.

In characterizing the sensitivity, it is crucial to take account of measurement backaction. With larger applied power or improved matching, the capacitance sensitivity can be improved at the price of a larger voltage drop across the device, potentially disturbing the state being measured. The figure of merit is therefore not simply $\SC$ but the product $\SC V_0$, where $V_0$ is the RMS excitation voltage at the device (see Supplementary Information). This is plotted in Fig.~\ref{Fig3}(b), with $V_0$ calculated from the carrier power using the circuit parameters from Fig.~\ref{Fig2}. For all data in Fig.~\ref{Fig3}, the carrier power $P_1$ at port 1 was set to $-29$~dBm and near perfect matching $V_0=117\pm54~\mathrm{\mu V_{RMS}}$, i.e the maximum bias applied was approximately 170~$\mathrm{\mu V_{RMS}}$. The figure of merit $\SC V_0$ is minimized at the same circuit tuning as $\SC$, confirming that the optimal configuration of the circuit is indeed close to perfect matching. Note that minimizing $\SC V_0$ is not achieved by minimizing $\CS$, but by setting $\CS$ to achieve perfect matching.

\begin{figure}
\includegraphics[width=\columnwidth]{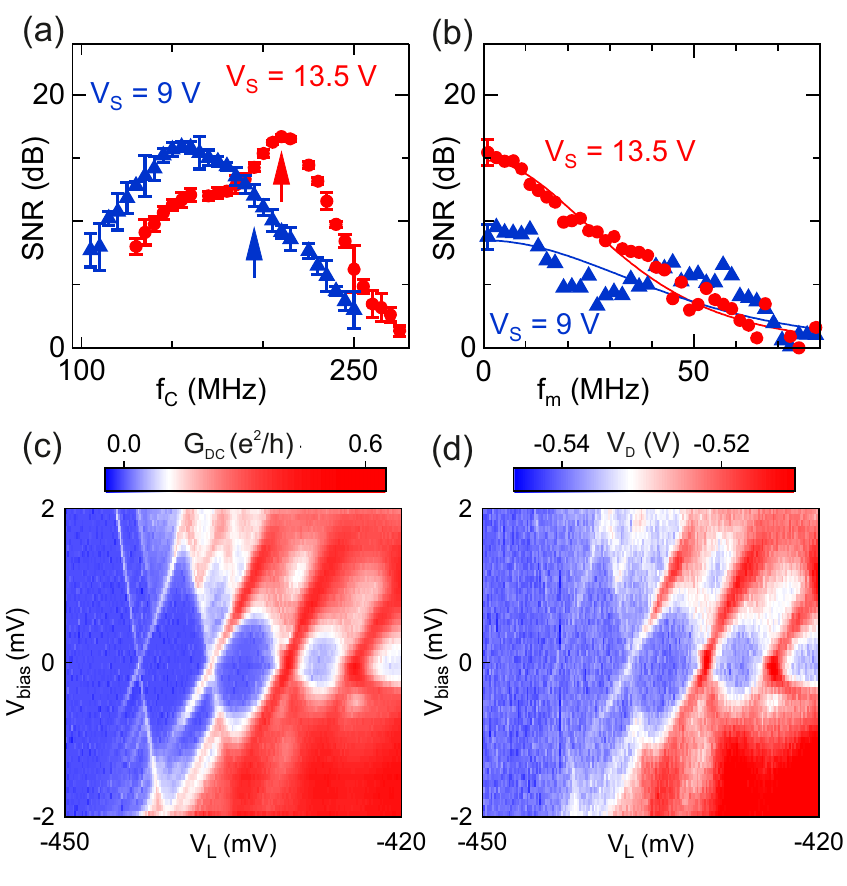}
\caption{
(a), (b) Comparison of SNR close to perfect matching ($\VS = 13.5$~V) and away from perfect matching ($\VS = 9$~V), with modulation applied to the gate voltage $\VL$. SNR is plotted against carrier frequency in (a) with $\fM=1.1$~MHz, and against modulation frequency in (b) with $\fC=\fo$ (marked with arrows in (a)). For both datasets, $\Delta f=10$~Hz, the modulation amplitude is 0.48~mV$_\mathrm{RMS}$, and $\TMC=1$~K. The applied power in (a) is adjusted at each frequency and voltage setting so as not to broaden the Coulomb peaks (see Supplementary Information); in (b), the power is fixed at the value chosen for $\fC=\fo$ ($P_{1}\approx-31$~dBm both at 9~V and 13.5~V). As seen from (a), the SNR is maximized near perfect matching and for $\fC\approx \fo$, although a resonance of the cryostat at low frequency enhances SNR around $\fC=160$~MHz. From Lorentzian fits (curves in (b)), the 3~dB readout bandwidth can be extracted. Error bars indicate variation between different data sets in (a) and the uncertainty in the noise level in (b); for clarity, only a single error bar is marked in (b).
(c) Conductance through the quantum dot as a function of $\Vbias$ and $\VL$, measured at $\TMC=20$~mK and showing Coulomb blockade diamonds. (d) Demodulated voltage $\VD$, measured simultaneously with (c) with $\VS=13.5$~V ($\fC=210.75$~MHz). The applied power, $P_{1}=-40$~dBm, did not broaden the Coulomb peaks at $\TMC=20$~mK.
\label{Fig4}
}
\end{figure}

\section{Measuring the quantum dot impedance}
We now turn to measurements of the quantum dot. First, we confirm that the impedance of the device itself can be measured with good sensitivity and bandwidth. Gate voltages were adjusted to the flank of one Coulomb peak at a point of maximum transconductance. With a modulation voltage now applied to a gate, Fig.~\ref{Fig4}(a) shows the sideband SNR as a function of \fC for two different varactor settings. Again, the perfect matching condition (still corresponding to \VS=13.5~V) yields a bigger SNR. Figure~\ref{Fig4}(b) shows SNR as a function of $\fM$, from which the readout bandwidth can be extracted; this is found to be 34~MHz at \VS = 9~V and 18~MHz at \VS=13.5~V. These data confirm that the readout bandwidth is set by the $Q$ factor of the circuit, and that the tradeoff between bandwidth and sensitivity can be tuned via a varactor.

Next, the stability diagram of the quantum dot is measured (see Supplementary Information). With the circuit cooled to $\TMC=20$~mK, simultaneous measurements of the DC transport conductance and the demodulated signal $\VD$ are shown as a function of $\VL$ and $\Vbias$ (Fig.~\ref{Fig4}(c) and (d)). Coulomb blockade diamonds are evident in both Fig.4(c) and (d). The similarity between these plots shows that changes in the quantum dot impedance are well captured in reflectometry. Although the setup is optimized for capacitance sensitivity measurements, we can also operate our device as a single-electron transistor and estimate its charge sensitivity. The charge sensitivity was calculated from the measured SNR at a given bandwidth and the charge modulation, estimated from the modulation amplitude and the Coulomb peak spacing~\cite{Schoelkopf_1998}, in an analogous expression to the one used for $\SC$.
We obtain $\sim 1650~\mu \text{e}/\sqrt{\text{Hz}}$ with a maximum $V_0$ of $144~\mathrm{\mu V_{RMS}}$. For Si transistors, the state-of-the-art value of 37~$\mu e\mathrm{Hz}^{-1/2}$ was achieved with an applied voltage to the RF gate of 0.5~mV~\cite{Gonzalez_2015}. Our diminished charge sensitivity reflects the smaller RF bias, the smaller lever arm and the life time broadening of the Coulomb peaks with respect to ref.~\cite{Gonzalez_2015}. 

\begin{figure}
\includegraphics[width=\columnwidth]{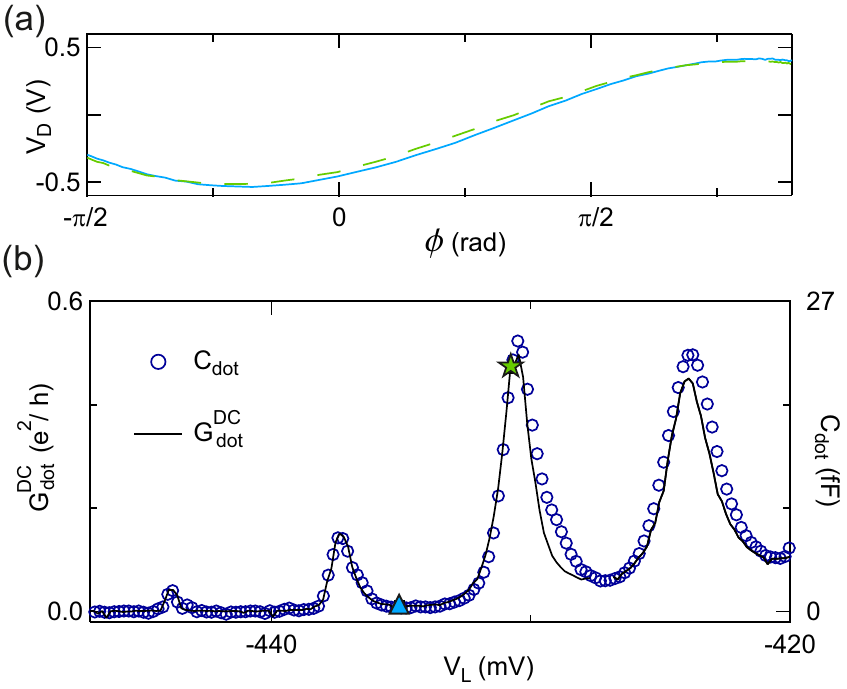}
\caption{\label{Fig5}
(a)
Demodulated response of the circuit as a function of $\phi$ for $\VL$ set on (green dashed) and off (blue solid) a Coulomb peak with $\Vbias=0$. Both amplitude and phase of $\VD$ are different, indicating a dissipative as well as a dispersive contribution on-peak.
(b)
Imaginary contribution to the device impedance as a function of $\VL$ over a series of Coulomb peaks. Line: measured DC conductance from Fig.~\ref{Fig4}(c). A gap in the line shows that we have accounted for a switch event that occurred after the data in Fig.~\ref{Fig4}(c) was taken. Symbols: capacitance extracted from curves as in (a) and a circuit model of the matching network. $C_{\text{dot}}$ varies proportionally to $\GdotDC $ as a function of $\VL$ .}
\end{figure}

In the data of Fig.~\ref{Fig4}(d), the demodulated signal is sensitive to both conductance and capacitance of the quantum device. To isolate the capacitance $\Cdot$, we measure $\VD$ as a function of the phase shift $\phi$ applied in the demodulation circuit. Figure~\ref{Fig5}(a) shows traces measured on and off a Coulomb peak, showing the phase shift associated with the quantum capacitance. To extract $\Cdot$ from the measured phase shift, is not sufficient simply to assume they are proportional, because changes in the quantum dot conductance also lead to a phase shift; however, using the measured DC conductance $\GdotDC$ within our circuit model, it is possible to calculate $\Cdot$ (see Supplementary Information). Figure~\ref{Fig5}(b) shows $\Cdot$ calculated at each gate voltage over a series of Coulomb peaks, together with $\GdotDC$ measured at the same settings.

It is evident that the quantum dot capacitance is proportional to the conductance. This reflects the fact that both quantities are proportional to the density of states of the quantum dot (see Supplementary Information). This contrasts with previous measurements where the tunnel barriers are more opaque and non-proportionality between conductance and capacitance can be observed ~\cite{Gabelli_2006, Chorley_2012}. This is the case, for instance, when the quantum dot dynamics is dominated by the quantum charge relaxation effect \cite{Buttiker_1993}, evidenced in RF conductance measurements. This rich phenomenology~\cite{Nigg_2006, Hamamoto_2010, Filippone_2011, Cottet_2015} can be explored with our setup.

\section{Discussion}

These sensitive measurements of quantum dot impedance are promising for readout of singlet-triplet spin qubits in a double quantum dot. Using quantum capacitance for readout of a singlet-triplet obviates the need for a charge sensor~\cite{Petersson_2010}, which is attractive for scalable two-dimensional architectures. However, although the theoretical sensitivity of this technique~\cite{Johansson_2006} allows for single-shot readout in a few microseconds, practical sensitivities have been found to be well below this, in part because of poor impedance matching.

Estimating the difference in quantum capacitance~\cite{Petersson_2010} between qubit states as $\sim 10$~fF, our measured sensitivity $\SC=1.6~\mathrm{aF}/\sqrt{\mathrm{Hz}}$ would at first sight indicate single-shot readout with unit SNR in integration time $T_\mathrm{meas} \sim 13$~ns. Crucially, this sensitivity is achieved with a maximum bias $V_0\approx170~\mathrm{\mu V_{RMS}}$, which is smaller than the typical singlet-triplet splitting in a qubit device~\cite{Petta_2005}, and therefore does not induce charge relaxation in the triplet manifold. However, this calculation does not take into account the fact that the quantum capacitance peaks in a narrow bias range near zero detuning. The single-shot readout time should instead be estimated by comparing the product $\SC V_0$, which characterizes the sensitivity to charge induced on the source electrode, with the actual charge  $\lambda e$ induced by electron tunneling, where $\lambda$ is the lever arm. Taking $\lambda=0.3$ from Fig.~\ref{Fig4} and the mean value of $\SC V_0$ close to perfect matching, we find that unit SNR requires $T_\mathrm{meas} \sim 64~\mu s$. Since this is about twice the singlet-triplet qubit relaxation time~\cite{Barthel_2009} in GaAs, further improvements will be required to achieve single-shot readout. Our approach can be improved by optimizing remaining geometric capacitance in the circuit, by using superconducting inductors to increase the quality factor~\cite{Churchill2012,Colless_2013}, and by using a superconducting amplifier with drastically reduced noise temperature ~\cite{Stehlik_2015}.

\section{Acknowledgements}
We acknowledge discussions with M. Shea and J. Medford.
Research at Oxford University was supported by EPSRC (EP/J015067/1), DSTL, the European Union, and the Royal Academy of Engineering. This publication was also made possible through support from Templeton World Charity Foundation. The opinions expressed in this publication are those of the authors and do not necessarily reflect the views of Templeton World Charity Foundation.

\vfill
%


\newpage

\end{document}


\setcounter{figure}{0}
\renewcommand{\thefigure}{S\arabic{figure}}
\setcounter{equation}{0}
\renewcommand{\theequation}{S\arabic{equation}}
\onecolumngrid
\parbox[c]{\textwidth}{\protect \centering \Large \MakeUppercase{Supplementary Information}}
\rule{\textwidth}{1pt}
\twocolumngrid

\title{Sensitive radio-frequency measurements of a quantum dot by tuning to perfect impedance matching}

\author{N.~Ares}
\affiliation{Department of Materials, University of Oxford, Parks Road, Oxford OX1 3PH, United Kingdom}

\author{F.J.~Schupp }
\affiliation{Department of Materials, University of Oxford, Parks Road, Oxford OX1 3PH, United Kingdom}

\author{A.~Mavalankar}
\affiliation{Department of Materials, University of Oxford, Parks Road, Oxford OX1 3PH, United Kingdom}

\author{G.~Rogers}
\affiliation{Department of Materials, University of Oxford, Parks Road, Oxford OX1 3PH, United Kingdom}

\author{J.~Griffiths}
\affiliation{Cavendish Laboratory, J. J. Thomson Avenue, Cambridge CB3 0HE, United Kingdom}

\author{G.A.C.~Jones}
\affiliation{Cavendish Laboratory, J. J. Thomson Avenue, Cambridge CB3 0HE, United Kingdom}

\author{I.~Farrer}
\affiliation{Cavendish Laboratory, J. J. Thomson Avenue, Cambridge CB3 0HE, United Kingdom}

\author{D.A.~Ritchie}
\affiliation{Cavendish Laboratory, J. J. Thomson Avenue, Cambridge CB3 0HE, United Kingdom}

\author{C.G.~Smith}
\affiliation{Cavendish Laboratory, J. J. Thomson Avenue, Cambridge CB3 0HE, United Kingdom}

\author{A.~Cottet}
\affiliation{Laboratoire Pierre Aigrain, Ecole Normale Sup\'{e}rieure-PSL Research University,
CNRS, Universit\'{e} Pierre et Marie Curie-Sorbonne Universit\'{e}s,
Universit\'{e} Paris Diderot-Sorbonne Paris Cit\'{e},
24 rue Lhomond, 75231 Paris Cedex 05, France}

\author{G.A.D.~Briggs}
\affiliation{Department of Materials, University of Oxford, Parks Road, Oxford OX1 3PH, United Kingdom}

\author{E.A.~Laird}
\affiliation{Department of Materials, University of Oxford, Parks Road, Oxford OX1 3PH, United Kingdom}

\date{\today{}}
\maketitle

\section{Circuit simulation of the matching network}

\begin{figure}[b]
\includegraphics[width=\columnwidth]{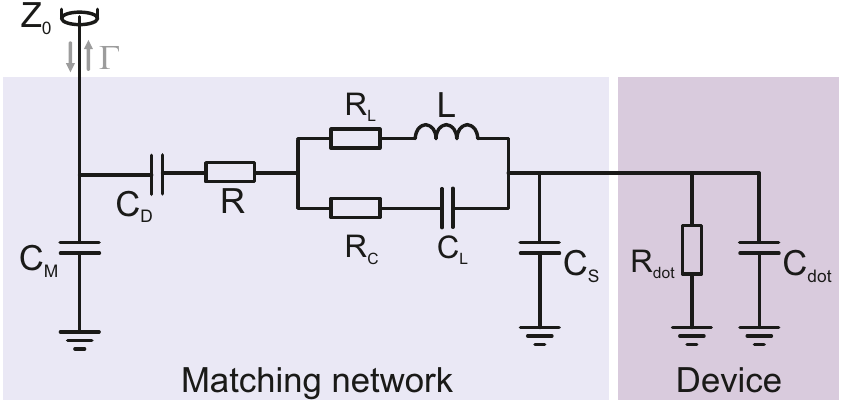}
\caption{\label{S1}
Circuit model of the matching network and device. Capacitors $\Cshunt$, $\Cseries$ and $\CS$ are taken as simple lumped elements, including any parasitic capacitances in parallel. Elements  $R_{\text{L}}$, $R_{\text{C}}$ and $C_{\text{L}}$ model parasitic contributions to the impedance of the inductor $L$. The effective resistance $R$ models other losses in the circuit. The quantum dot is modelled by the combination of $R_{\text{dot}}=1/\GdotRF$ and $C_{\text{dot}}$.
}
\end{figure}

The fits in Fig.~2 are obtained using the circuit model shown in Fig.~\ref{S1}. The three capacitors $\CS$, $\Cseries$ and $\Cshunt$ are taken as simple lumped elements, each incorporating any parasitic capacitance in parallel with the varactor. The inductor is modelled as a network of elements as shown, which simulate its self-resonances and losses. Other losses in the circuit are modelled by an effective resistance $R$. The device under test is taken as a parallel $RC$ circuit.

The reflection coefficient $\Gamma$ is then equal to
\begin{equation*}
\Gamma(\fC) = \frac{\Ztot(\fC)-\ZO}{\Ztot(\fC)+\ZO},
\label{eq:s11}
\end{equation*} 
where $\Ztot$ is the total impedance from the circuit's input port and $Z_0=50~\Omega$ is the line impedance. We relate the measured transmission $S_{21}$ to $\Gamma$ by assuming a constant overall insertion loss $A$, incorporating  attenuation in the cryostat lines, the coupling of the directional coupler, and the gain of the amplifier, such that 
\begin{equation}
|S_{21}(\fC)|=A|\Gamma(\fC)|.
\label{eq:s21}
\end{equation} 

\begin{figure}[b]
\includegraphics[width=\columnwidth]{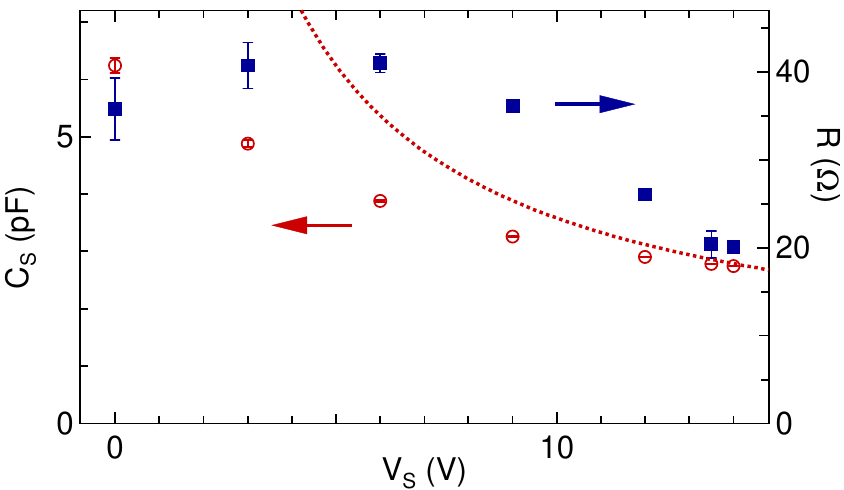}
\caption{\label{S0}
Values of $\CS$ and $R$ at $\TMC=1$~K, obtained from fits as in Fig.~2(a). Dashed line is the datasheet value of $\CS$ at room temperature including a parasitic capacitance of 0.7~pF. The comparatively small fitted values of $\CS$ indicate that some of the dopants in the varactor have frozen out.
}
\end{figure}

Fitting to Eq.~(\ref{eq:s21}), we take $\Cseries=87$~pF from the known component value; take $L=223$~nH, $R_{\text{L}}=3.15 \times 10^{-4}~\Omega \times \sqrt{\fC~[\mathrm{Hz}]}$, R$_{\text{C}}=25~\Omega$ and C$_{\text{L}}=0.082$~pF from the datasheet of the inductor; and assume $\Rdot = 1~\mathrm{G}\Omega$ and $\Cdot=1~\mathrm{aF}$ for a pinched-off device. Fit parameters are then $A$, $\Cshunt$, $R$, and $\CS$. From the fit at $\VS=13.5$~V, we obtain $A=-27.6\pm0.3$~dB and $\Cshunt=14.5\pm0.9$~pF. For other fits, we hold these values constant;  extracted values for  $\CS$ and $R$ at each voltage are plotted in Fig.~\ref{S0}.

Below 1~K this GaAs junction varactor shows a delayed response to the tuning voltage, making acoustic-frequency modulation unreliable at $T_\mathrm{MC}=20$~mK. Nevertheless, it continues to act as a radio-frequency capacitor, although with diminished and temperature-dependent tuning range. For this reason, exact impedance matching could not be achieved at 20~mK on this cooldown; that would require increasing the geometrical contribution to $\CS$ or decreasing that to $\Cshunt$.

\section{Circuit sensitivity}
In this section we give further details of how the sensitivities in Fig.~3 are measured and simulated.

\subsection{Measuring the sensitivity}\label{deltaC}
\begin{figure}
\includegraphics[width=\columnwidth]{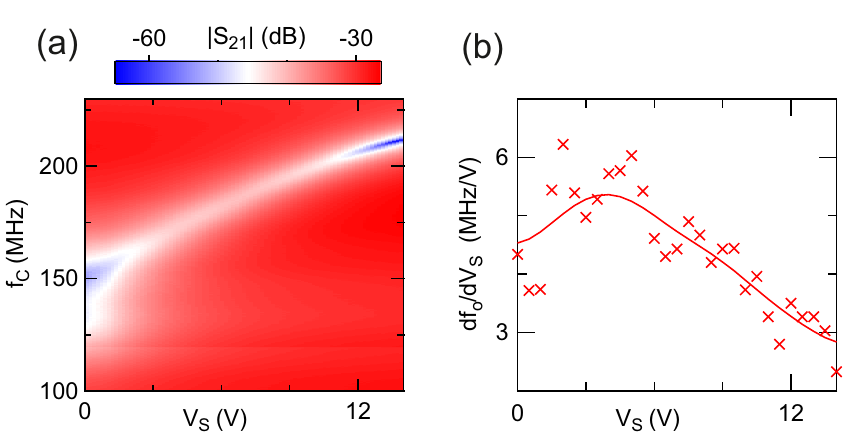}
\caption{\label{S2}
(a) Response of the circuit as a function of $\VS$ and $\fC$. (b) Points: Numerical derivative ($d\fo/d\VS$) as a function of~$\VS$. Line: Smoothed data used in Eq.~(\ref{eq:deltaC*}) to calculate $\delta \CS$. 
}
\end{figure}

For the sensitivity measurement in Fig.~3, we determine the root-mean-square (RMS) capacitance modulation $\delta \CS$ as follows. From a measurement of $|S_{21}|$ as a function of $\fC$ and $\VS$ (Fig.~\ref{S2}(a)), we extract the resonance frequency $\fo(\VS)$, defined as the location of the minimum at each voltage. Approximating that
\begin{equation}
\fo(\VS)\approx\frac{1}{2\pi\sqrt{L\CS(\VS)}},
\label{eq:f0approx}
\end{equation}
we relate $\delta \CS$ to the RMS varactor modulation voltage $\VM$ by:
\begin{eqnarray}
\delta \CS 	&=& \left|\frac{d\CS}{d \VS}\right |\VM \nonumber \\
			&=& \frac{\VM}{2 \pi ^2 L\fo^{3}} \left|\frac{d\fo}{d\VS}\right|,
\label{eq:deltaC*}
\end{eqnarray}
where $d\fo/d\VS$ is taken as a smoothed numerical derivative of the measured $\fo(\VS)$~(Fig.~\ref{S2}(b)).

Although~Eq.~(\ref{eq:f0approx}) applies strictly only for a simple $LC$ resonator, we have confirmed numerically that this procedure gives a good approximation for $d\CS/d\fo$ in our circuit model.

\subsection{Simulating the sensitivity}
Taking the amplitude of the incident signal as $V^0_\mathrm{in}$, the reflected signal from the matching network is:
\begin{equation*}
V_\mathrm{r}(t) = V^0_\mathrm{in}~\mathrm{Re}(\Gamma e^{i\omega t}).
\end{equation*}
In response to a change $\delta\CS$ in the capacitance, the reflection coefficient changes by $\delta \Gamma$, leading to a change in reflected voltage:
\begin{equation*}
\delta V_\mathrm{r}(t) = V^0_\mathrm{in}~\mathrm{Re}(|\delta \Gamma| e^{i (\omega t + \mathrm{arg}(\delta \Gamma))}),
\end{equation*}
giving for the variance in $V_\mathrm{r}$
\begin{eqnarray*}
\langle \delta V_\mathrm{r}^2 \rangle 	&=& \frac{(V^0_\mathrm{in})^2}{2} \langle |\delta \Gamma|^2\rangle\\
						&=& \frac{(V^0_\mathrm{in})^2}{2}\left|\frac{d\Gamma}{d\CS}\right|^2 \langle \delta \CS^2\rangle.
\end{eqnarray*}
The sensitivity is defined as the root-mean-square $\delta \CS$ per unit bandwidth for which $\langle\delta V_\mathrm{r}^2\rangle$ becomes equal to the noise fluctuation $S_V^2 \Delta f$:
\begin{equation*}
S_C = \frac{\sqrt{2}}{\left|\frac{d\Gamma}{d\CS}\right|V^0_\mathrm{in}} S_V.
\end{equation*}
Assuming that the system noise is dominated by the amplifier noise temperature $T_\mathrm{N}$ and that there are no losses between the matching circuit and the amplifier input, we obtain:
\begin{equation}
S_C = \frac{\sqrt{2kT_\mathrm{N}\ZO}}{\left|\frac{d\Gamma}{d\CS}\right|V^0_\mathrm{in}},
\label{eq:SC}
\end{equation}
where $k$ is Boltzmann's constant. This is the formula used to simulate $S_C$ in Fig.~3.

In this simulation, we took the amplifier noise from the manufacturer's specification, giving $T_\mathrm{N}=3.7$~K. We calculated $\left|\frac{d\Gamma}{d\CS}\right|$ numerically using the same circuit model as above, taking parameters from fits as in Fig.~2. The signal level to the matching circuit, $V^0_\mathrm{in}$, is in principle known from the applied power $P_1=-29$~dBm and the insertion loss in the injection line. However, we find that the fits in Fig.~2 yield an overall insertion loss ($A=-27.6\pm0.3$~dB) that is greater than the sum of the fixed attenuators on the injection line (-31~dB), the coupling of the directional coupler (-20~dB), and the specified amplifier gain (+32~dB). This implies a distributed additional insertion loss of $\sim -8.6$~dB, with a corresponding uncertainty in the value of  $V^0_\mathrm{in}$. For the simulations in Fig.~3, we assumed this contribution was distributed equally before and after the matching network, but the possibility of unequal distribution dominates the error bars in simulated $S_C$ and $S_CV_0$.

\subsection{The figure of merit for readout bandwidth}
In the main paper, we stated that the figure of merit for a dispersive qubit readout circuit is not the capacitance sensitivity $\SC$, but rather the product $\SC V_0$. In this section, we justify this statement.
For any transition being measured, such as the inter-dot transition in a double-dot qubit, the quantum capacitance peaks in a window about zero detuning. Although Eq.~(\ref{eq:SC}) predicts an improving capacitance sensitivity with increasing incident signal, once the detuning excitation becomes larger than the peak width, over most of the detuning cycle the device is configured to have zero capacitance. The benefit of the larger drive voltage is therefore lost.

To calculate the bandwidth with which the inter-dot transition can be resolved, we note that the average capacitance over the RF cycle is
\begin{equation*}
\overline{C} = \frac{\Delta q}{\Delta V}
\end{equation*}
where $\Delta V = 2\sqrt{2}V_0$ is the peak-to-peak voltage on the source electrode and $\Delta q$ is the change in electrode charge over the cycle. The maximum value of $\Delta q$ is achieved when $V_0$ is set much larger than the peak width, and is given by $\Delta q = \lambda e$, where $\lambda$ is the lever arm. The readout bandwidth $\Delta f$ for unit SNR then satisfies:
\begin{eqnarray}
\sqrt{\Delta f} &=& \frac{\overline{C}}{\SC} = \frac{\Delta q}{2\sqrt{2}S_C V_0} \nonumber \\
		 &=& \frac{\lambda e}{2\sqrt{2}S_C V_0}.
		 \label{eq:SCV0}
\end{eqnarray}
For given lever arm, the figure of merit is therefore given by $S_C V_0$, which from Eq.~(\ref{eq:SC}) is fixed by the circuit parameters independent of the incident power. For the optimum tuning of this circuit, we have $S_C V_0 = 1.2 \times 10^{-3} e / \sqrt{\mathrm{Hz}}$, so for $\lambda=0.3$ we could achieve unit SNR for single-shot readout in bandwidth $\Delta f = 7.8$~kHz, or an integration time of 64~$\mu$s.

\section{Carrier signal power}

\begin{figure}[b]
\includegraphics[width=\columnwidth]{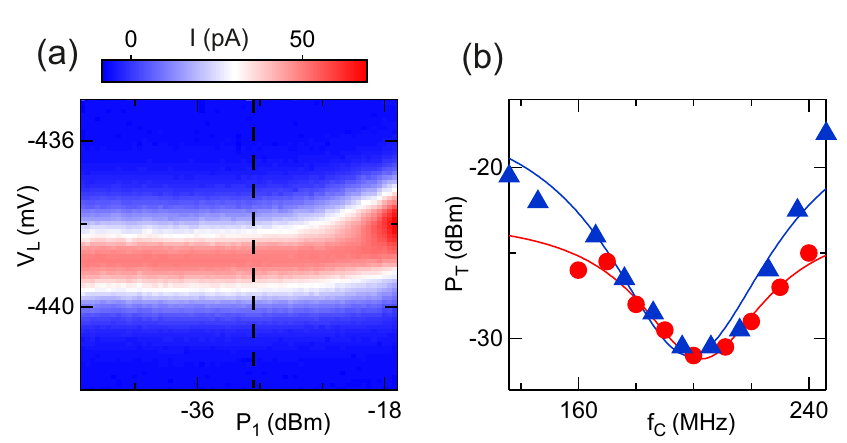}
\caption{\label{Sp}
(a) Current as a function of $\VL$ and $P_{1}$, showing the Coulomb peak at which the SNR measurements in Fig.~4(a) and (b) were performed. For these data, $\VS=13.5$~V, $\fC=211$~MHz, and  $\Vbias=100~\mu$V. The dashed line indicates the threshold carrier power $P_{\text{T}}$; for $P_{1}<P_{\text{T}}$ no detectable broadening of the Coulomb peak is observed. (b) Measured $P_\mathrm{T}$ (points) as a function of carrier frequency for $\VS=9$~V (triangles) and $\VS=13.5$~V (circles). Solid lines are Lorentzian fits from which values of $P_{1}$ were chosen for the SNR measurements in Fig.~4(a).
}
\end{figure}

The carrier power in Fig.~4(a) was chosen separately at each frequency $f_C$ to avoid broadening the Coulomb peak. Fig.~\ref{Sp}(a) shows this peak broadening for increasing $P_{1}$ for a typical combination of $\fC$ and $\VS$. For each such combination, a threshold power $P_\mathrm{T}$ is extracted, defined as the largest power for which no peak broadening could be detected. Figure~\ref{Sp}(b) shows how this threshold power depends on $\fC$ for two different $\VS$ settings. Each dataset is fitted to a Lorentzian to define a function $P_\mathrm{T}(\fC, \VS)$. This is the carrier power chosen in Fig.~4(a).

\section{Measurement of the quantum dot capacitance}

To determine the device capacitance $\Cdot$ as in Fig.~5(b), we begin by measuring $|S_{21}(\fC)|$ with $\VL$ set so that the device is pinched off. In this situation, we assume that $\Cdot=0$. Fitting as in Fig.~2, we obtain $\CS= 2.673\pm0.005$~pF, $\Cshunt= 15.0\pm0.8$~pF, $R=11.7\pm0.9~\Omega$ and $A= -24.9\pm0.2$~dB. The differences from the values at 1~K indicate that parasitic losses are reduced at 20~mK.

Given these circuit parameters, the phase of the reflected signal for other $\VL$ settings is determined entirely by $\Cdot$ and $\Rdot$. From the measured phase shift as a function of $\VL$, obtained by fitting curves similar to Fig.~5a, and assuming $\Rdot=1/\GdotDC$, we numerically evaluate $\Cdot$ to obtain the values plotted in Fig.~5(b).

\section{Proportionality between quantum dot capacitance and conductance}\label{sec:theory}
This section presents a simple model which explains the relationship between $\GdotDC$, $\GdotRF$ and $\Cdot$. We consider a quantum dot coupled to left and right leads by tunnel rates $\GammaL$ and $\GammaR$ respectively, and assume the low-temperature limit $k\TMC/\hbar \ll \GammaR, \GammaL$. The linear DC conductance can then be written
\begin{equation}
\GdotDC=\frac{2e^2}{h}\frac{\GammaL \GammaR}{\GammaL + \GammaR}~\rho(\epsilonF,\VL)
\label{eq:DCconductance}
\end{equation}
where $\rho$ is the quantum dot density of states, $\epsilonF$ is the the Fermi level and $V_{\text{L}}$ the gate voltage that allows us to control the electrochemical potential of the quantum dot~\cite{Meir_1991}. 

We now consider a time-dependent voltage $V(t)=\sqrt{2}V_0 \cos \omega t$ applied to the left lead. The current that flows in response to this voltage is not in general~\cite{Gabelli_2006,Chorley_2012} determined by Eq.~(\ref{eq:DCconductance}). However, so long as $\omega \ll \GammaL, \GammaR$, electron tunnelling occurs instantaneously on the timescale of the oscillating voltage and there is no additional dissipation due to the time dependence. From the width of the smallest Coulomb peak in Fig.~5(b), we deduce $\GammaL+\GammaR \sim 30$~GHz, and therefore this assumption is justified. The RF conductance is therefore the same as the DC conductance:
\begin{equation*}
\GdotRF=\GdotDC.
\end{equation*}

A capacitance arises because the quantum dot charges and discharges in response to the RF voltage. The charge on the dot is
\begin{equation*}
Q(t) = Q_0 + e\kappa \rho(\epsilonF, \VL) V(t)
\end{equation*}
where $Q_0$ is the charge with no bias, $e$ is the elementary charge, and $\kappa$ is the occupation probability of a state in the transport window. To supply this charge, the current from the left lead is
\begin{eqnarray*}
I(t)	 &=& \beta\dot{Q}(t)\\
	 &=& e\beta\kappa \rho(\epsilonF, \VL) \dot{V}(t)
\end{eqnarray*}
with $\beta$ a constant which parameterizes how much of the charge tunnels from the left lead, as well as how plasmonic screening currents triggered by tunneling events are distributed in the circuit \cite{Blanter_2000}. We identify this current as due to a capacitance
\begin{equation*}
\Cdot=e\beta\kappa \rho(\epsilonF, \VL).
\end{equation*}
Comparing with Eq.~(\ref{eq:DCconductance}), we see that 
\begin{equation*}
\Cdot = \beta \kappa \frac{\GammaL+\GammaR}{\GammaL\GammaR} \GdotDC
\end{equation*}
This is consistent with Fig.~5, provided that the proportionality constant stays approximately equal as a function of $\VL$. From the fact that the proportionality constant is similar over adjacent Coulomb peaks, we deduce that $\beta \kappa$ increases with increasing $\VL$ at approximately the same rate as $(\GammaL+\GammaR)/\GammaL\GammaR$ decreases.

\vfill